# Research on Intelligent Traffic Control Methods at Intersections Based on Game Theory

Huisheng Wang, Yuejiang Li, H. Vicky Zhao

**Abstract:** Based on game theory and dynamic Level-$k$ model, this paper establishes an intelligent traffic control method for intersections, studies the influence of multi-agent vehicle joint decision-making and group behavior disturbance on system state. The simulation results show that this method has a good performance when there are more vehicles or emergency vehicles have higher priority.

## I. Introduction

Intelligent traffic system (ITS) uses communication network and artificial intelligence technology to collect different levels of traffic information and decision-making behavior, establish dynamic model of traffic system, and carry out short-term and long-term management and control according to the information and model, so as to improve transportation efficiency, alleviate traffic congestion, reduce traffic accidents, realize the harmony of people, vehicles and roads, and provide better service for traffic participants.

At the intersection controlled by traffic lights, the decision-making of vehicles only depends on the signal timing in most cases, that is to say, when there is no accident, vehicles follow the traffic rule of "red for stop and green for go". When the signal timing is set reasonably, the intersection can usually achieve the optimal situation.

In the intersection without traffic signal control, each vehicle can make any decision to maximize the individual utility, which forms a game. But the result of the game is not always satisfactory. Sometimes it may fall into a low-level equilibrium, which is reflected in traffic congestion or accidents at intersections. Therefore, how to make rules to make the game as close as possible to the optimal solution of the system is a problem to be solved.

The game of intersection is infinite, but the payoff function is only related to the results of the following several times, and the discount factor is small. This kind of foresight can be analyzed by dynamic Level-$k$ model. Different from straight road traffic control, the state space of intersection is more diversified, and when the number of vehicles is large, it is easy to cause traffic congestion, making intersection control more complex.

## II. Literature review

In recent years, intelligent transportation and automatic driving have been increasingly concerned by industry and academia. Intelligent traffic control technology, i.e. information processing and analysis system, has various algorithms, such as ant colony algorithm, Dijkstra algorithm and A* algorithm for shortest path optimization based on trajectory planning; and Yolo algorithm based on road recognition and target detection to calculate the speed and density of vehicles, judge the length of time for drivers to reach the destination, and avoid congestion. [6][7]

For the intelligent traffic control of intersections, many researches focus on the design of traffic signal timing, and then guide the decision-making of vehicles. At present, reinforcement learning, which is very popular, is widely used in signal timing. The state space of signal timing is less, and only four signal phases can be set at the simplest intersection. Guo used the method of combining deep reinforcement learning with traffic signal control to transform the traffic signal control problem into a reinforcement learning problem interacting with the intersection on a discrete time step, taking the waiting time of the intersection as the objective function. The results show that the waiting time of vehicles under different saturation flow rates has different degrees of improvement, which verifies the effectiveness of the algorithm. [8] Chen applied deep reinforcement learning to intelligent control of traffic signals, which can obtain good traffic signal control strategies and effectively reduce vehicle stay time, vehicle delay and traffic congestion. Among them, the average driving time of vehicles was reduced by 34.1%, the average queue length was shortened by 19.4%, and the average waiting time was reduced by 24.0%. [9] Zhu proposed an adaptive and coordinated intelligent traffic signal control system, which can detect the traffic condition of the road in real time, and regulate the traffic light duration according to the traffic flow of each road section, and establish the system control between the signal lights, so as to ensure the shortest waiting time of vehicles for the red light, and realize the maximum dredging effect of the road. [10]

The complexity of the game theory model is related to the dimension of the state space (the number of branches of the extended subgame) and the dimension of the strategy space. Compared with intersections, straight roads are studied earlier using game theory models. The vehicle status on a straight road can be described by a two-dimensional matrix. The strategy space only involves two possibilities of longitudinal acceleration (forward direction) and deflection acceleration (lane change), so it is easier to study. Li studied online traffic control based on game theory and proposed a method to simulate driver interaction behavior under given traffic conditions. [5] In addition, combined with the support vector machine model with high comprehensive prediction performance, the recognition accuracy of lane changing behavior can reach 92.97% , surpassing other models and much higher than the regular model. Evolutionary game theory has also been applied to lane change decision model. [11]

In the game theory control of intersections, Li proposed a game theory method of multi-stage and interactive decision-making model of vehicles at unsignalized intersections. This paper mainly studies the situation of traffic accidents when two vehicles are driving in opposite directions without intervention. The dynamic Level-$k$ model is used to model the time-varying multi-step vehicle interaction at unsignalized intersections. The results show that compared with different types of drivers, the traffic control system has the ability to solve the intersection conflict. [4]

In this paper, we will modify and improve the above model to study the situation of multi-agent vehicle joint decision-making. On this basis, we will model its characteristics and possible behaviors, describe its decision-making process and the interaction between various actors, study the influence of group behavior on individual behavior, and analyze the disturbance on the system state. According to the above research results, effective system governance rules and incentive mechanism are formulated to guide user behavior and make the system achieve the desired state. At the same time, the dynamic Level-$k$ model is applied to analyze the decision-making behavior and system performance of emergency vehicles.

### III. Model Construction
1. **Game Theory and Dynamic Level-$k$ Model**

Game theory is the method of studying the phenomena of confrontation or competition. Each party participating has different interests, so it is necessary to consider various possible action plans of the opponent and choose the most beneficial plan. [1]

A complete game model contains at least the following three basic elements: players, strategy space and payoff function. Use $I = \{1, 2, ..., n\}$ to represent the collection of $n$ players. Each player $i \in I$ has its own set of strategies $S_i$. All players' strategies formed a policy group, called a situation, $\boldsymbol{s} = (s_1, s_2, ..., s_n)$. It is recorded $S = \{\boldsymbol{s}\}$ as the collection of all situations. When a situation $\boldsymbol{s}$ occurs, a winning value or loss value $H_i(\boldsymbol{s})$ is specified for player $i$, defined as the payoff function. When the above three elements are determined, a game model is also determined.

Game theory has two basic assumptions:

(1) **Rationality**: Player $i \in I$ concerns with maximizing its own profit: $\max_{\boldsymbol{s} \text{ or } s_i} H_i(\boldsymbol{s})$.

(2) **Common Knowledge**: Not only all players are rational, but also all players know the other players are rational, all players are aware of all the players know the other players are rational ...... and so on.

According to players action order, games can be divided into simultaneous game and sequential game. Simultaneous game requires that players $i, j \in I$ make decisions $s_i, s_j$ at the same time or although the two players do not make decisions at the same time, they do not know the other's choice when making decisions. In the sequential game, the actions of the players $i, j$ have a sequence, and the later player $j$ can observe the decisions chosen by the first player $i$.

According to the information the players have, games can be divided into complete information game and incomplete information game. In the complete information game, player $i \in I$ not only know its own payoff $H_i(\boldsymbol{s})$, but also aware of other players' payoff, $H_k(\boldsymbol{s}), k \in I, k \neq i$.

Compared with decision theory, the important feature of game theory is the interdependence of decision-making. The optimal strategy $s_i$ of player $i \in I$ depends on the strategy selection of other decision makers $k \in I, k \neq i$. In some cases, the optimal strategy $s_i^*$ of the player $i$ may not depend on the strategy selection of other players $k$, which is called the dominant strategy. Nash equilibrium is a general concept of complete information simultaneous game solution. Many games do not have dominant strategy equilibrium, but there is Nash equilibrium.

For sequential games, if a strategy combination is a Nash equilibrium in every subgame, then the strategy combination is called subgame perfect Nash equilibrium (SPNE). The SPNE requires that the decision must be optimal on the information set composed of individual nodes, so the Nash equilibrium selected by using the perfect concept of subgame must be optimal. The method of solving the limited perfect information game with extended expression is reverse induction. [2]

However, for many well-known games, such as repeated prisoner's dilemma and the chain store market entry game and so on, the unique solution obtained by the inverse induction method was rarely selected by real subjects (about 6%). Generally, there are two stylized facts concerning the violations of backward induction: limited induction and time unraveling. In the simple game with fewer subgames, players seldom violate the reverse induction, and the behavior of players converges in time to the reverse induction. [3]

The dynamic Level-$k$ model solves the contradiction between the above two situations. The premise of the model is that players have different reasoning levels, and the level $k$ is used to represent the degree of foresight. Players with a level of 0 make their own optimal decisions, called instinctive decisions, regardless of interactions with others. Players with level 1 consider such interactions in the decision-making process and assume that the level of others is 0. In other words, except for level 0, each player assumes that the level of all other players is one level lower than that of itself and makes instinctive decisions; the player predicts his behavior and the game evolution caused by its behavior, and makes his own decision as the optimal response to this development. Similarly, the player at level $k$ assumes that all other players are level $k-1$, predicts based on this assumption and responds. [4]

**2. Intersection Game Model**

The following will introduce the principle of the intersection game model in accordance with the sequence of the establishment process.

(1) **From the decision-making behavior of individual vehicles, the traffic problem at the intersection forms a game.**
(2) **As the vehicle makes decisions all the time, it is an infinite complete information game.**
(3) **Vehicles have different queuing priorities, making decisions in a certain order, which changes the simultaneous game into a sequential game.**
(4) **Because the state of the vehicle is different from that of the last decision, it is a sequential game with different structure.**
(5) **Because vehicles only focus on the utility of several times, the farther the utility is, the less influence it has on current decision-making.**

**IV. Conclusion**

Based on the game theory and dynamic Level-$k$ model, this paper establishes the intelligent traffic control method of intersection, studies the situation of multi-agent vehicle joint decision-making, and the influence of group behavior on individual behavior and system state, and analyzes the influence of disturbance on system state. The dynamic Level-$k$ model is also applied to analyze the decision-making behavior and system performance of emergency vehicles.

The simulation results show that under the fixed model, the increase of the probability of occurrence of random factors, the increase of the number of input vehicles, and the decrease of vehicle time interval will cause vehicle congestion and overall driving speed reduction. Under the fixed random factors, vehicle number and vehicle time interval, the Level-1 game model is better than the Level-0 model. The performance of the higher Level-$k$ model is close to that of the level-1 model.

According to the above results, within a certain range, the greater the maximum speed and the smaller the minimum safety distance, the shorter the average driving time. However, the way to change the preference weight will not stimulate the individual behavior and improve the system performance. In the unsignalized intersection, the maximum speed and the minimum safety distance can be set to formulate effective system governance rules and incentive mechanism to guide drivers' behavior.